\documentclass{epl}

\newcommand{\lest}{{l}}

\title{Spatial correlations in the relaxation of the Kob-Andersen model}

\author{Enzo Marinari\inst{1} and Estelle Pitard\inst{2}}
\institute{
\inst{1}Dipartimento di Fisica, SMC and UdR1 of INFM, INFN, 
Universit\`a di  Roma {\em La Sapienza}, P.le Aldo Moro 2, 00185 Roma, Italy.\\
\inst{2}Laboratoire des Verres (CNRS-UMR n$^o$5587), CC69, Universit\'{e}
Montpellier 2, 34095 Montpellier Cedex 5, France.}

\pacs{64.70.Pf}{Glass transitions}
\pacs{05.50.+q}{Lattice theory and statistics}
\pacs{61.20.Lc}{Time-dependent properties of liquid structure; relaxation}

\begin{document}

\maketitle

\begin{abstract}
We describe spatio-temporal correlations and heterogeneities in a
kinetically constrained glassy model, the Kob-Andersen model.  
The kinetic constraints of the model alone induce the
existence of dynamic correlation lengths, that increase as the
density $\rho$ increases, in a way compatible with a
double-exponential law.  We characterize in detail the trapping time
correlation length, the cooperativity length, and the distribution of
persistent clusters of particles. This last quantity is related to the
typical 
size of blocked clusters that slow down the dynamics for
a given density.
\end{abstract}


Glassy dynamics in kinetically constrained systems has been a subject
of wide interest during the last years \cite{RITSOL}. It indeed
includes a large class of models of particles constrained by dynamical
rules where the relaxation is slowed down when the density of
particles increases or the temperature decreases, sharing similarities (though with a number of
differences) with spin-glass models or glass models \cite{CUGLIA} when
the temperature is decreased.  In the former class, slowing down of
the dynamics can be understood as a consequence of steric effects,
that make certain moves of particles impossible because of an
effective high local density, putting strong constraints on the number
of allowed paths between configurations \cite{WHIGAR}; whereas for the
latter class, frustration and/or disorder in the interactions produces
a slow time relaxation.  In real glasses both effects are likely to be
present. While most studies on the glass state have focused for a long
time on the latter class of models, more attention is paid to the fact
that the former mechanism is also of importance, especially since the
works of Harrowell \cite{HARROW} on spatio-temporal heterogeneities.
These heterogeneities are also of great interest for experimentalists, since they have been
identified in several
glassy systems \cite{EDIGER}.

In this paper we will focus on the numerical study of the Kob-Andersen
(KA) model \cite{KOBAND}. In this 3-dimensional model of lattice gas
on a cubic lattice, a particle can hop to one of its empty nearest
neighbor sites only if it has strictly less than $m$ occupied
nearest-neighbor sites, and if after the hop it will also have less
than $m$ occupied nearest-neighbor sites (which ensures detailed
balance). In the original study of the KA model, the value of $m$ is
chosen to be $m=4$, which is a compromise between free diffusion of
particles with excluded volume ($m=6$) and a truly non-ergodic model
(for $m\leq 3$ small cubes of occupied sites are blocked for
ever). The relaxation is hence a diffusive process with additional
steric constraints. The energy landscape is trivial in the sense that
there is no interaction between particles, hence all configurations
are thermodynamically equally probable.  However the original
numerical study of the KA model showed features similar to those seen
in glassy systems such as the arrest of the diffusion constant at some
density $\rho_c$ smaller than 1 ($\rho_c \simeq 0.881$ in
\cite{KOBAND}), and slow stretched-exponential relaxation of the
intermediate scattering function; though, it was already pointed out
by the authors that the Mode-Coupling predictions for glasses
are actually not
valid for this model. Moreover, the dynamics is time translationally
invariant (there is no aging), because all configurations are
equivalent. One way to induce aging is to do density crunches (i.e
quenches to higher densities) in a system where a reservoir of
particles is added, as studied in \cite{KUPESE}, or to add explicit
two-body interactions between the particles.

Recently the question of finite-size effects in the KA model, already
addressed in \cite{KOBAND}, came back, related to the still unsolved
issue of a true dynamical transition in the thermodynamic limit.  It
is now established that in the thermodynamic limit and in finite
dimension, no dynamical arrest occurs at a density smaller that one
\cite{TONINE,BIFITO,ADLER}; whereas in a mean-field version of the
model on a Bethe lattice, such a dynamical transition at $\rho_c < 1$
occurs.  More precisely, the critical density in three dimensions
depends on the size $L$ of the system like $1- \rho_c(L) =C/(\ln(\ln
L))$, where $C$ is a constant, which makes the convergence to $1$ very
slow and unreachable by numerical simulations. In other words, at
fixed density $\rho$, in order to eliminate finite size effects, one
has to study system sizes much larger than the spacing between mobile
particles $\Xi(\rho) \propto \exp(\exp(C/(1-\rho)))$.

Finally the KA model has been shown to share an important feature with
the dynamics of mean-field spin-glass models and supercooled liquids
\cite{EDIGER,FRAPAR,DFGP,LACEVI,FRMUPA}: the dynamical susceptibility $\chi(t)$ shows a
pronounced maximum at an intermediate time $t^{\star}$,
this maximum is interpreted as the time where heterogeneities are the
strongest in the system, and increases as the density increases
(analog to a decrease of temperature in other models), hinting at the
existence of a growing correlation length.  Because of its purely
constrained nature (no energy barriers, no dynamical transition) the
origin of these spatio-temporal heterogeneities in the KA model comes
merely from steric effects \cite{TONINE,CHAGAR}.

%
In this study we give a description of the spatio-temporal
correlations and heterogeneities that characterize the dynamics of the
KA model.  First, we characterize the distribution of trapping
times of particles (for recent work on different systems see for example
\cite{DOLHEU,BODERE}) and its spatial correlations.  Then we describe
the behavior of the four-point spatio-temporal correlation functions
in real space, giving  a determination of a
cooperative length-scale not provided a priori by $\chi(t)$, in an approach complementary to
 \cite{BERTHIER}. Finally we
study the dynamics of the KA model in terms of the distribution of
persistent clusters (this is a new approach to the problem), and show
how spatial heterogeneity and cooperativity can be identified in this
way. We show that the different lengths identified in this study
behave in the way predicted by Toninelli, Biroli and Fisher
\cite{TONINE,BIFITO,ADLER} as a function of the density.

\section{Trapping times}

\begin{figure}
\includegraphics[width=10cm, height=4cm]{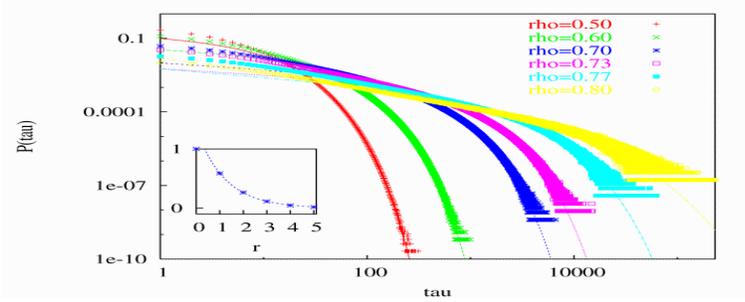}
\caption{$P(\tau)$ versus $\tau$ in for different values
of $\rho$.  The smooth curves are for the stretched
exponential best fits for the integrated probability.  In the inset
$C_{\rho}(r)$ for $\rho=0.77$.  
\protect\label{F-PTAU}}
\end{figure}

We have computed the distribution of trapping times (or 
persistence times) $P(\tau)$.  Let $\tau_i$ be the time after which
the particle starting from site $i$ first moves to one of its
nearest-neighbor site: we calculate the probability for $\tau_i$ to
have the value $\tau$, and we average over all particles and over a
large number of initial configurations of the system. We have computed
$P(\tau)$ for different densities (ranging from $0.5$ to $0.8$, see
figure \ref{F-PTAU}).  Our fits are consistent with a stretched
exponential behavior for the integrated probability distribution,
$\int_{\tau}^{\infty} dt\ P(t)=\exp{\left\{ -(\tau/\tau_{trap}
)^{\beta} \right\}}$, where $\beta$ decreases as the density increases
from $0.80$ ($\rho=0.5$) to $0.45$ ($\rho=0.8$).  Such stretched
exponential behavior also characterizes $P(\tau)$ in KCM, in
Lennard-Jones systems and in experiments (see \cite{BERGAR} and
references therein). In particular, in the East model, a temperature dependent stretching exponent for
$P(\tau)$ is observed \cite{BUGAR}.

%
The site dependent trapping times defined above do not need to be
homogeneous in space. One can ask whether the KA
dynamics induces a spatial correlation length among them. We have
quantified this aspect by computing spatial correlations:
$C_{\rho}(r)\equiv\frac{1}{N} \sum_{\vec{a}} (\langle \tau_{\vec{a} }
\tau_{\vec{a}+\vec{r}}\rangle - \langle \tau_{\vec a}\rangle \langle
\tau_{\vec{a}+\vec{r}}\rangle)$, where $\langle \cdots \rangle$ stands
for an average over initial configurations.  $\lest_c(\rho)$ is a {\em
dynamical coherence length} such that for distances larger than
$\lest_c(\rho)$ the trapping times are uncorrelated.  An exponential
fit, where $C_{\rho}(r)\sim e^{-r/\lest_c(\rho)}$, done for $r$ going
from $1$ to $5$ works well.  Although in the density range $0.7$-$0.8$
$\lest_c(\rho)$ is small and of the order of one or two lattice
spacings, the growth of $\lest_c(\rho)$ for increasing $\rho$ (from
0.7 to 0.77) can be fitted by a double-exponential law $\lest_c(\rho)
\simeq 0.17 \exp(\exp(0.16/(1-\rho)))$.

%
The finding of this stretched exponential behavior leads to consider
the relaxation of correlation functions. 
Ref. \cite{SIMMOU} has analyzed stretched
exponential relaxations in glassy systems, looking at it as a
consequence of a renewal process
for a single effective particle, where the distribution of renewal
times is itself a stretched exponential, leading to stretched-exponential relaxation
 with the same exponent for
the time correlations. 
It is not obvious whether the cooperative dynamics of the KA model could be mapped onto 
such a simple renewal
process with a distribution of times equal to the distribution
of persistence times $P(\tau)$. To test this, we define the overlap function as
$q(t)\equiv\frac{1}{N\rho(1-\rho)} \sum_i (n_i(t)n_i(0) -\rho^2)$,
where $n_i(t)=0 {\rm \ or \ } 1$ is the occupation number of site $i$
at time $t$.  We have computed $\langle q(t)\rangle$ as an average
over initial configurations.  Even if in the asymptotic time regime
$\langle q(t)\rangle$ should decay as a power law \cite{SPOHN}, we do
not see any sign of this behavior \footnote{We have checked that even
in the unconstrained lattice gas for high density values we are not
able to observe a simple power law decay.}: it is in fact very well
fitted by a stretched exponential, with a stretching exponent that
depends on $\rho$, $\langle q(t)\rangle \sim \exp{ (-(t/t_{relax})^{
\gamma} )}$.  $\gamma$ decreases from $0.8$ to $0.35$ for $\rho$
increasing from $0.5$ to $0.8$. The exact values of $\gamma$ are not
identical to the ones of $\beta$, but they are very similar: in the
limits of the statistical and systematic error, that we estimate to be
smaller than $0.05$,  we find the values compatible.  A possible real
discrepancy of the two exponents would be a signal of the fact that
the growth of spatial correlations with density
(although $l_c(\rho)$ is always small in our simulations)
plays an important role, and that the KA model can not be reduced to a
unique  effective renewal process. Instead,
one has to take into account the spatial correlations of the dynamics.

\section{Spatio-temporal correlation functions}

The dynamic susceptibility $\chi(t)\equiv N \left(\langle
q^2(t)\rangle - \langle q(t)\rangle^2\right)$ has been one of the main
evidences presented in \cite{FRMUPA} stressing the relation of the KA model
to
liquids and spin glasses: as a function of time it shows a maximum at
$t^{\star}$, where the sensitivity of the system is maximal.  We have
repeated with larger precision, for a range of densities, the
measurements of \cite{FRMUPA}, finding compatible results.  Both the
value of the maximum of $\chi(t)$ and $t^{\star}$ as a function of
$\rho$ are very well fitted by the double-exponential scaling form of
\cite{BIFITO} (see also Figure \ref{F-FIG6}).  At $t=t^{\star}$ (which is of the same order of
magnitude as the relaxation time) the statistical heterogeneity of the
system is maximal. We show now that it is also related to spatial
cooperativity and heterogeneity.

\begin{figure}
\includegraphics[width=4cm, height=10cm,angle=270]{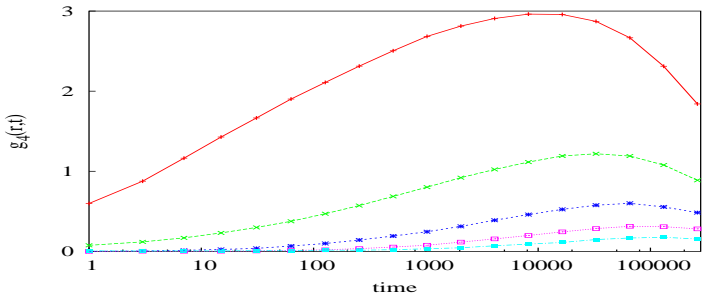}
\caption{$g_4(r,t)$ for $\rho=0.85$ and $L=24$ as a function of time.
From top to bottom, $r=1,2,3,4,5$.
\protect\label{F-G4RT}}
\end{figure}

We define and measure a space-dependent susceptibility $g_4(\vec{r},t)$,
which generalizes $\chi(t)$ \cite{LACEVI}:
\begin{equation}
g_4(r,t) \equiv \left( N\rho^2 (1-\rho)^2 \right)^{-1} 
\sum_{|{\vec{r}}_i - {\vec{r}}_j|=r}
\left(\langle n_i(t)n_i(0)n_j(t) n_j(0) \rangle - \langle n_i(t)n_i(0)
\rangle \langle n_j(t)n_j(0)\rangle\right)\ . 
\end{equation}
%
Here we only take the spatial
separation on a single coordinate axis, so that $r$
is at the same time  an Euclidean and a Manhattan distance. 
We show $g_4(\vec{r},t)$ as a function of time in figure \ref{F-G4RT}:
for all values of $r$ it shows a maximum.  The position of the maximum
is of the same order of magnitude as $t^{\star}$ but shifts to larger
times as $r$ increases. The value of the maximum is decreasing with
$r$.

\begin{figure}
\includegraphics[width=4cm, height=10cm,angle=270]{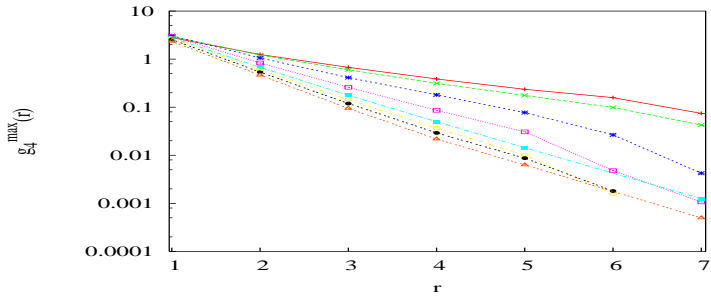}
\caption{$g_4^{max}(r)$ as a function of the distance $r$.  From top
to bottom, $\rho=0.86, 0.85, 0.83, 0.8, 0.77, 0.75, 0.73, 0.7$.
\protect\label{F-G4RMAX}}
\end{figure}

The function $g_4(\vec{r},t)$ gives information on how the decorrelation between
times $0$ and $t$ at site $i$ is related to the decorrelation of site
$j$ between the same times.  The decrease of $g_4^{max}(r)$ (in figure
\ref{F-G4RMAX}) enables us to determine a cooperativity length above
which the decorrelating events are not correlated (or not
``cooperative'').  We fit to the form $g_4^{max}(r) = \frac {g(\rho)}
{r^{\alpha(\rho)}} \exp^{-r/\xi(\rho)}$.  The fit works well enough:
for $\rho$ in the range $0.7-0.86$ we find values of $\alpha(\rho)$
and $\xi(\rho)$ ranging from $0.22$ to $0.64$, and from $0.70$ to
$2.75$ respectively. Again the cooperativity length $\xi(\rho)$ is
very well fitted by a double exponential form:
$\xi(\rho)\simeq 0.34 \exp(\exp 0.16/(1-\rho))$. We have determined a
time scale and a cooperativity scale $\xi(\rho)$ that diverge as
$\rho\to 1$. 
Note that the behaviour of $g_4^{max}(r)$ is more reminiscent of a strong glass of
the type of the 3d Fredricksen-Andersen model,
than of a fragile glass (see \cite{NEWREF}), although the KA model is also fragile,
but with a conserved order parameter.

\section{Connected clusters of frozen particles}

\begin{figure}
\includegraphics[width=4cm, height=12cm,angle=270]{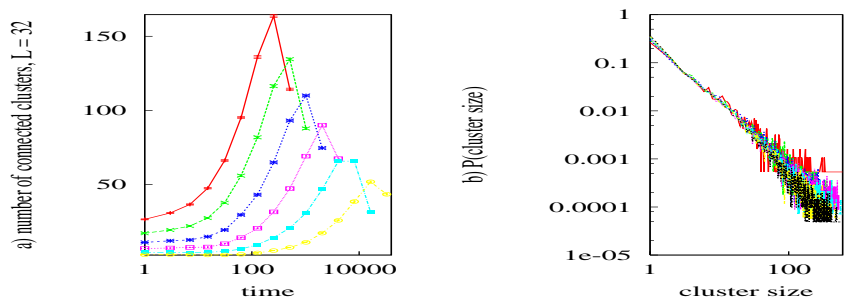}
\caption{For $\rho$ going from 0.7 to 0.8:
a) number of persistent connected clusters (lower densities on the
  left, higher densities on the right); 
b) probability distribution of the number of particles in
persistent clusters at $t_{cluster}(\rho)$ for all densities: one can observe the collapse of the curves. 
\protect\label{F-FRACTAL}}
\end{figure}

In exploring length and time scales of KA we have also focused on the
analysis of the connected clusters of  blocked particles: we
reconstruct the clusters of nearest-neighbor
particles that have not moved at $t$ from
their initial position.  We have analyzed them by computing their
number and the probability distribution of clusters of size $n$,
$P(n,t)$.
We show in figure
\ref{F-FRACTAL}.a the number of such blocked clusters, that turns out
to be maximum at a time $t_{cluster}$. 
We plot with different lines data at different density values.
This time scale, close to
$\tau_{trap}$ and $t_{relax}$, has again a behavior compatible with
\cite{BIFITO}.

At small times the probability $P(n,t)$ is mainly composed of two
distinct populations: large blocked clusters that dominate the rate of
relaxation and small blocked clusters.  As time passes some particles
move, and the large initial blocked clusters decrease in size, giving
rise to smaller clusters.  At an intermediate timescale (which is very
close to $t_{cluster}$) the two populations of blocked clusters
merge. We show in figure \ref{F-FRACTAL}.b $P(n,t_{cluster})$ for
different density values.  Here the behavior looks as a power law
($\sim \frac{1}{n^{\mu}}$, where $\mu \sim 1.3$) and
one can observe a collapse of the curves for all densities; the origin of this
unique exponent is unclear to us.  At very long times the system
relaxes completely: the persistent clusters only consist of a few
particles, that will eventually fly.  For $t\sim t_{cluster}$ the
initial structure of clusters has been destroyed on small and large
length scales: at this stage spatial cooperativity is maximum, and one
can see that the particles not only move from the borders of blocked
clusters, but also propagate inside, breaking big clusters into
smaller ones of significantly different sizes.

%
We have studied the moments $\lambda(\rho)=\langle
\lambda\rangle_{P^*}$ and $\sigma(\rho)=\sqrt{\langle
\lambda^2\rangle_{P^*}-\langle \lambda\rangle_{P^*}^2}$, where
$\lambda\equiv n^{1/3}$, $n$ is the cluster size and expectation
values are taken over the probability distribution $P^*$ at
$t_{cluster}$. $\lambda(\rho)$ is 
the typical size of the blocked clusters that slow down the dynamics, and can
hence be seen as a (maybe rough) determination of the length
$\Xi(\rho)$, defined in \cite{BIFITO} as the spacing between mobile
regions. A fit actually gives $\lambda(\rho)\simeq 0.83 \exp (\exp
(0.12/(1-\rho)))$.  The fluctuations measured by $\sigma(\rho)$ obey
the same exponential law in $\rho$, but reach higher values than
$\lambda(\rho)$.

\section{Conclusions}


\begin{figure}
\includegraphics[width=5cm, height=10cm,angle=270]{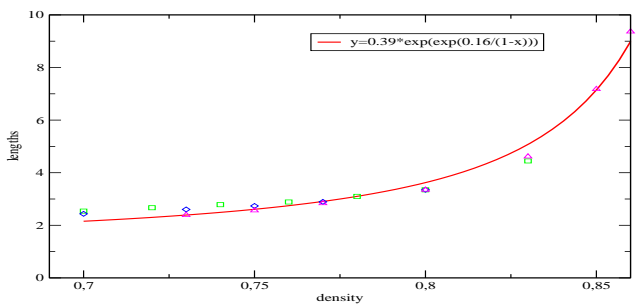}
\caption{The three lengths, $\lambda(\rho)$ (squares), 
$1.78 \lest_c(\rho)$  (diamonds), and $3.41 \xi(\rho)$ (triangles)
versus $\rho$, together with the best double-exponential fit.
\protect\label{F-FIG5}}
\end{figure}

We have shown that dynamical geometrical constraints in the KA model
are able to
induce spatial correlations through different characteristic length
scales, that all diverge in the same way as $\rho\longrightarrow 1$.
We show in figure \ref{F-FIG5} the behaviour of $\xi(\rho)$,
$\lest_c(\rho)$, and $\lambda(\rho)$.  They 
can be rescaled by a constant, and are shown together with the fit
$y \sim \exp(\exp 0.16/(1-\rho))$.  

\begin{figure}
\includegraphics[width=10cm, height=5cm,angle=0]{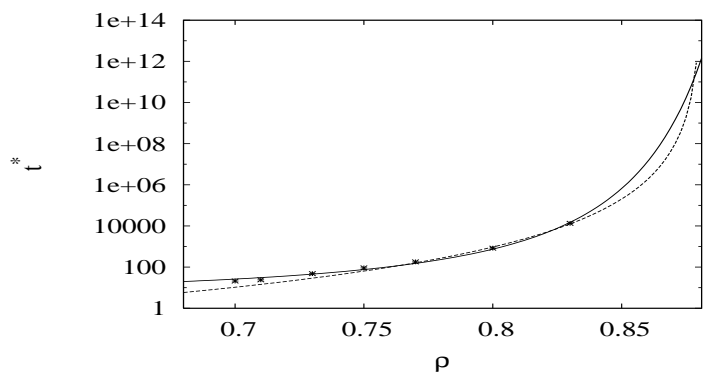}
\caption{$t^*$ versus $\rho$. The continuous line is for the best fit
  for a double-exponential singularity at $\rho=1$, the dashed line is for the best power
  fit (with $\rho_c=0.881$).
\protect\label{F-FIG6}}
\end{figure}

We have already noticed that also time scales (that grow in our
allowed range far more than length scales, giving a far better way
of discriminating between different behaviors) suggest a scaling law
compatible with a double-exponential singularity at $\rho=1$. We show in particular
in figure
\ref{F-FIG6} the time $t^*$ defined from the maximum of $\chi(t)$ versus
$\rho$: the fit with a double-exponential is surely preferred over the best fit
to a power divergence at $\rho_c=0.881$ (note that the $y$ scale is
logarithmic, and the power fit is far from the data).

In our study, $\lest_c(\rho)$ measures the correlation between sites of the
persistence times.  The coherence length $\xi(\rho)$ deduced from
measurements of four-point correlation functions is the typical length
of cooperative events for a given density.  The typical length between
mobile particles that contribute to the relaxation of the system
$\Xi(\rho)$ can be estimated via $\lambda(\rho)$, the average size of
persistent clusters up to $t_{cluster}$, the time of maximum
cooperativity.  Moreover, the spatial structure contains even more
complexity, since the distribution of this length obeys a power-law,
which still remains to be understood.  An analytical solution of the
KA model is lacking, and would give a more precise understanding of
the relaxation laws, the persistence properties, the existence of a
dynamical exponent $z$, and the distributions of cluster sizes,
probably to be linked with an analogy with percolation problems
\cite{KOBAND, BIFITO,ADLER}.

Finally, we want to stress that the coherence length $\xi(\rho)$ is a
priori independent of any growing length-scale that is believed to
exist in a system without time-translational invariance dynamics
(aging).  In the version of the KA model studied here, there is no
aging due to the flat energy landscape. However, an aging version
would be able to show how $\xi(\rho, t_w)$ evolves with the age $t_w$.
In particular it would be very interesting to study its behavior and
that of $t^{\star}(\rho,t_w)$ in this case and compare it to
experimental data in aging jammed systems \cite{CIP}.

This work has been supported by EEC contracts HPRN-CT-2002-00307
(DYGLAGEMEM) and HPRN-CT-2002-00319 (STIPCO), and by the ESF SPHINX
network.  EP thanks UMR 5825 (Montpellier) where this work was
initiated, and the Physics Department at {\em La Sapienza}.  We thank
L. Berthier, J-P. Bouchaud, S. Franz, C. Godr\`eche, R. Monasson,
V. Van Kerrebroeck and especially G. Biroli, W. Kob and C. Toninelli
for discussions.

\bibliographystyle{apsrev}

\begin{thebibliography}{99}

\bibitem{RITSOL}
F. Ritort and P. Sollich,
Adv. in Phys. {\bf 52}, 219 (2003). 

\bibitem{CUGLIA} 
L. F. Cugliandolo, 
in {\em Slow Relaxations and Non-Equi\-li\-brium Dynamics in Condensed 
Matter}, 
edited by J.-L. Barrat et al.
(Springer, Berlin 2003).

\bibitem{WHIGAR}
S. Whitelam and J. P. Garrahan, 
J. Phys. Chem. B {\bf 108}, 6611 (2004).

\bibitem{HARROW}
S. Butler and P. Harrowell, 
J. Chem. Phys. {\bf 95}, 4454 (1991);
P. Harrowell.
Phys. Rev. E {\bf 48}, 4359 (1993).


\bibitem{EDIGER}
M.D. Ediger,
Annu. Rev. Phys. Chem. {\bf 51}, 99 (2000).


\bibitem{KOBAND}
W. Kob and H. C. Andersen, 
Phys. Rev. E {\bf 48}, 4364 (1993).

\bibitem{KUPESE}
J. Kurchan, L. Peliti and M. Sellitto,
Europhys. Lett. {\bf 39}, 365 (1997).

\bibitem{BERGAR}
L. Berthier and J. P. Garrahan, 
J. Chem. Phys. {\bf 119}, 4367 (2003);
L. Berthier and J. P. Garrahan, 
Phys. Rev. E {\bf 68}, 041201 (2003).









\bibitem{TONINE}
C. Toninelli,
PhD Thesis, Universit\`a di Roma {\em La Sa\-pien\-za},
October 2003, unpublished.

\bibitem{BIFITO}
C. Toninelli, G. Biroli and D. Fisher, 
Phys. Rev. Lett. {\bf 92}, 185504 (2004);
C. Toninelli and G. Biroli,
cond-mat/0402314. 

\bibitem{ADLER}
J. Adler, 
Physica A {\bf 171} (1991) 435.


\bibitem{FRAPAR}
S. Franz and G. Parisi, 
J. Phys C {\bf 12}, 6335 (2000).

\bibitem{DFGP}
C. Donati, S. Franz, G. Parisi and S. Glotzer, 
J. Non-Cryst. Sol. {\bf 307}, 215 (2002).

\bibitem{LACEVI} 
N. Lacevic, F. Starr, T. Schroeder, S. Glotzer,
J. Chem. Phys. {\bf 119}, 7372 (2003);  
L. Berthier,
Phys. Rev. Lett. {\bf 91} 055701 (2003).

\bibitem{FRMUPA}
S. Franz, R. Mulet and G. Parisi,
Phys. Rev. E {\bf 65}, 021506 (2002).

\bibitem{CHAGAR}
J. P. Garrahan and D. Chandler,
Phys. Rev. Lett. {\bf 89}, 035704 (2002).

\bibitem{DOLHEU}
B. Doliwa and A. Heuer,
Phys. Rev. E {\bf 67}, 030501(R) (2003).

\bibitem{BODERE}
R. A. Denny, D. R. Reichman and J.-P. Bouchaud,
Phys. Rev. Lett. {\bf 90}, 025503 (2003).


\bibitem{BERTHIER}
L. Berthier, Phys. Rev. Lett. {\bf 91}, 055701 (2003).

\bibitem{BUGAR}
A. Buhot, J.P. Garrahan, Phys. Rev. E. {\bf 64},  021505 (2001).



\bibitem{SIMMOU}
S. I. Simdyankin and N. Mousseau,
Phys. Rev. E {\bf 68}, 041110 (2003).

\bibitem{SPOHN}
H. Spohn,
{\em Large Scale Dynamics of Interacting Particles}
(Springer, Berlin 1991).

\bibitem{CIP} 
L. Cipelletti, H. Bissig, V. Trappe, P. Ballestat,
S. Mazoyer, 
J. Phys. C {\bf 15} S257 (2003).

\bibitem{NEWREF}
S. Whitelam et al., cond-mat/0408694. L. Berthier, J.P. Garrahan, cond-mat/0410076.

\end{thebibliography}

\end{document}